\documentclass[acmsmall]{acmart}

\usepackage{xcolor}

\AtBeginDocument{%
  \providecommand\BibTeX{{%
    \normalfont B\kern-0.5em{\scshape i\kern-0.25em b}\kern-0.8em\TeX}}}

\begin{document}

\title{Reimagining Communities through Transnational Bengali Decolonial Discourse with YouTube Content Creators}

\author{Dipto Das}
\affiliation{%
  \department{Department of Computer Science}
  \institution{University of Toronto}
  \city{Toronto}
  \state{Ontario}
  \country{Canada}
}
\additionalaffiliation{
  \position{now graduated doctoral candidate}
  \department{Department of Information Science, }
  \institution{University of Colorado Boulder}
  \city{Boulder}
  \state{Colorado}
  \country{United States}
}
\email{diptodas@cs.toronto.edu}

\author{Dhwani Gandhi}
\affiliation{
  \institution{Independent Researcher}
  \state{Pennsylvania}
  \country{United States}
}
\email{dhwanigandhi97@gmail.com}

\author{Bryan Semaan}
\affiliation{
  \department{Department of Information Science}
  \institution{University of Colorado Boulder}
  \city{Boulder}
  \state{Colorado}
  \country{United States}
}
\email{bryan.semaan@colorado.edu}


\begin{abstract}
    Colonialism--the policies and practices wherein a foreign body imposes its ways of life on local communities--has historically impacted how collectives perceive themselves in relation to others. One way colonialism has impacted how people see themselves is through nationalism, where nationalism is often understood through shared language, culture, religion, and geopolitical borders. The way colonialism has shaped people's experiences with nationalism has shaped historical conflicts between members of different nation-states for a long time. While recent social computing research has studied how colonially marginalized people can engage in discourse to decolonize or re-imagine and reclaim themselves and their communities on their own terms--what is less understood is how technology can better support decolonial discourses in an effort to re-imagine nationalism. To understand this phenomenon, this research draws on a semi-structured interview study with YouTubers who make videos about culturally Bengali people whose lives were upended as a product of colonization and are now dispersed across Bangladesh, India, and Pakistan. This research seeks to understand people's motivations and strategies for engaging in video-mediated decolonial discourse in transnational contexts. We discuss how our work demonstrates the potential of the sociomateriality of decolonial discourse online and extends an invitation to foreground complexities of nationalism in social computing research.
\end{abstract}



\maketitle
\section{Introduction}
Nationalism creates a collective sense of identity--how individuals, despite their physical separation and limited personal interaction, see themselves as belonging to a large group~\cite{anderson2006imagined}, such as based on language, culture, religion, and even geopolitical borders. However, nationalism has a long and sordid history; a history where, when associated with modern nation-states, highlights a deep entanglement with the global history of colonialism~\cite{chatterjee1993nation, fanon2004wretched, anderson2006imagined}. Colonialism as an ideology, with colonization as its practice, refers to a system through which one country or group of people establishes control over another territory, often geographically distant, to establish settlements or colonies, exploit the latter's resources, and alter their political, economic, and cultural structures and identities~\cite{loomba2015colonialism} which later shaped local communities in terms of nation-states~\cite{chatterjee1993nation}. In this paper, we are interested in understanding the relationship between nationalism and colonialism and asking how sociotechnical systems like user-generated video-sharing platforms support decolonial discourse in transnational contexts--across nation-states with shared colonial history.

Colonial imposition of language, norms, laws, practices, institutions, social classification, and hierarchization schemes onto the colonized people~\cite{fanon2008black} and theft and vandalization of cultural sites and artifacts~\cite{van2004rethinking} prevented colonized people from developing an independent sense of identity and from understanding their own cultures. Thus, colonial values and categorizations profoundly impacted how previously colonized communities perceive themselves and their collective identities~\cite{fanon2008black}, fragmented native communities~\cite{chatterjee1993nation, chatterji2002bengal}, and introduced nationalistic sentiments and shaped nation-states~\cite{kaviraj2007making}. For example, through the colonial subjugation and partitions (see section~\ref{sec:research_site} for detailed background), the Bengali people and their identities have been fragmented by internal and external dislocations. Today, Bengalis find themselves in India, Pakistan, and Bangladesh. Their varied experience with colonization (e.g., West Bengalis experiencing only British rule but East Bengalis experiencing both British and Pakistani rules) differently shaped their conceptualization of nationalism (e.g., people in West Bengal identifying as Indian in contrast to people in East Bengal identifying as Pakistani till 1971 and as Bangladeshi after that).

In colonial and postcolonial societies, the process of undoing the transgenerational impact of colonization is called decolonization~\cite{fanon2004wretched, laenui2000processes}. While establishing sovereign nation-states is often viewed as the end goal of decolonization~\cite{laenui2000processes, fanon2004wretched}, the primary objectives of this process are reclaiming and reaffirming people's indigenous identities~\cite{fanon2008black} and reforming sociopolitical and economic structures in a way that reflects local values and perspectives~\cite{fanon2004wretched}. To attain those objectives, colonized communities must conceptualize their native culture and identity outside the colonial influences exercised through cultural imposition and territorial divisions in the form of nation-states. While the restoration of native culture and identity is often used as an umbrella concept in the literature as the decolonial objective~\cite{shaw2006encountering}, it abstracts the myriad historical internal and external dislocations and transnational fractures colonization caused to people's collective identities. While nationalism embeds fragmentary ways to imagine their communities and collective identities within people, different sociopolitical structures and institutions, such as mainstream media and education systems, materially bolster those ideas in postcolonial nation-states. Acknowledging the social psychology and materiality of decolonization as a political struggle~\cite{tuck2021decolonization, fanon2004wretched, laenui2000processes}, in this paper, we focus on decolonial discourse--discussions and practices that critically engage with and challenge colonial structures, ideologies, and power dynamics.

As institutions in postcolonial nation-states continue to reify colonial legacies through promoting dominant narratives and hyper-nationalism~\cite{sahoo2020mounting, lall2008educate} (elaborated in section~\ref{sec:lit_rev_institution_nationalism}), the processes of decolonization involve rediscovering cultural heritage, mourning about history, exploring decolonial possibilities, committing to reforming practices, and taking concrete actions~\cite{laenui2000processes}. Especially considering decolonization's socio-psychological aspects~\cite{fanon2004wretched, fanon2008black}, recovering cultural identity, understanding history, and examining colonial influence in regional geopolitics, economic hurdles, and social injustices become contributing factors to the decolonial discourse, which often takes shape and manifests through technology like online platforms~\cite{laenui2000processes, das2022collaborative}. Prior CSCW and social computing literature have highlighted how contemporary information and communication technologies (ICT) like online platforms can help marginalized communities--those who are pushed to the periphery of society, based on individual or multiple interconnected dimensions of identity~\cite{trudeau2011geographies}, to raise their voices and represent diverse opinions~\cite{das2022collaborative, rho2018fostering, saha2022towards}. Researchers have particularly explored how user-generated video-sharing online platforms support people in expressing and negotiating their cultural identities~\cite{milliken2008user-youtube, chen2023my} and participating in sociopolitical discussions~\cite{askanius2011online, askanius2014video}. Yet, what is less understood is how online platforms can support people in decolonizing nationalist systems of power and oppression, and this work aims to address this gap.

To answer this question, we focus particularly on Bengali identity and culture because its colonial history provides a great context for understanding the relationship between colonialism and nationalism and how that relationship shapes decolonial discourse online. Specifically, through a qualitative study realized through semi-structured interviews with participants from Bangladesh, India, and Pakistan, we investigated YouTubers' motivations and strategies for video-mediated transnational decolonial discourse. As collective identities shape contemporary ICT-mediated conversations on politics and international relationships, the goal and contribution of this social computing research is a deep understanding of people's use and practices around technologies (e.g., YouTube) in terms of their self-perceived nationalism. Moreover, challenging the use of umbrella terms like ``local", ``native," and ``Indigenous" communities in decolonial and postcolonial computing, building on Partha Chatterjee's work~\cite{chatterjee1993nation}, our paper highlights the fragments and nuances in imagined collectives among previously colonized communities. Our findings reveal how communities imagined in relation to language, religion, and postcolonial nation-states and institutionalized nationalism through different sociopolitical, media, and educational systems motivate YouTubers in decolonial discourse on YouTube. Moreover, our study explains their strategies, such as political explainers, storytelling, and YouTube journalism, that contribute to decolonial discourse. Our findings explain the double bind of nationalism--historically contributing significantly to anticolonial movements and simultaneously isolating regional and local identities bearing colonial divisions--and how it manifests in decolonial discourses on YouTube.

\section{Research Site: Bengali People's Experience with Colonization}\label{sec:research_site}
For a broader readership of this paper, in this section, we will briefly explain the Bengali people's complex historical experience with colonization. The Indian subcontinent (see Appendix~\ref{app:indian_subcontinent} for details about the geographic composition) is a site of prolonged British colonization. The Bengali (endonym: Bangali) people are one of the largest ethnolinguistic groups (approximately 259.89 million~\cite{index2020world, review2021world}) of this region, whose native language is Bengali (endonym: Bangla). They are native to the Bengal region that consists of present-day Bangladesh~\cite{bsb2022preliminary} and some Indian states like West Bengal, Assam, and Tripura~\cite{government2014report, india2014assam, state2015know}. This was the first British colonized region in the Indian subcontinent~\cite{das2022collaborative}. Throughout the colonial period, Bengal was the site of anticolonial movements (e.g., \textit{Swadeshi}) driven by Bengali nationalism~\cite{kaviraj2007making}. In 1947, when the British left, Bengal was used as a site of partition. West Bengal, which comprised a predominantly upper-caste Hindu majority, became a part of India, while East Bengal, which comprised a predominantly Muslim and lower-caste Hindu (e.g., \textit{namasudra}) majority became a part of Pakistan (named East Pakistan)~\cite{sen2018decline}. Since this partition and the formation of Pakistan were based on both political efforts and popular support, some characterize this arrangement as a ``partnership"~\cite{zaman1975emergence}. However, highlighting more on the realities and effects than the initial intent, many scholars argue that West Pakistan's (current Pakistan) myriad ways of economic extortion, political suppression, linguistic imposition, sociocultural marginalization, genocide, and rapes were concerted to oppress East Pakistan as a colony~\cite{ahmed1972struggle, das1978internal, nanda2019self}, and the assertion of Pakistan as a colonial force is not to weigh the impacts of British colonialism similarly or lightly. The Bengalis in East Pakistan fought for independence and formed Bangladesh in 1971~\cite{van2020history}. During 1947-1971 and soon after 1971, many Muslim Bengalis migrated to Pakistan for better economic opportunities~\cite{weinraub1973bhutto} and many Hindu Bengalis migrated to India fearing religious persecution~\cite{consulargeneral1971telegram}.

Because of such power dynamics throughout the history of Bangladesh, India, and Pakistan, their sociopolitical experience with colonization is complex and nuanced, especially in the case of the Bengali people. People in these three countries suffered under British colonization together until 1947. Then, the Bengali people in East Bengal experienced another 25 years (1947-1971) of foreign sociopolitical and economic rule and cultural imposition by Pakistan. Hence, the Bengali people's experiences with colonialism in West Bengal and East Bengal differ. Similarly, while in the case of British rule, people in Bangladesh, India, and Pakistan were colonized subjects, in the case of Pakistani rule, Bangladesh and Pakistan were essentially in a colonized-colonizer relationship (for a detailed rationale for this characterization in addition to the explanation above, please see~\cite{ahmed1972struggle, shafiqul1997impact}). Moreover, prolonged colonial effects, such as religious nationalism and geopolitical tension (e.g., control over Kashmir), adversely impacted the India-Pakistan relationship over the years~\cite{schofield2021kashmir}, while the Bangladesh-India relationship recently has some vital advancements (e.g., resolving border dispute, sharing waters of common rivers) with much room for progress~\cite{chakma2012bangladesh, hossain2016future, hossain1998bangladesh}.

Though prior works have looked at the identity expression of Bengali people in relation to colonization on online platforms~\cite{das2022collaborative, das2021jol}, researchers have not explored how different periods of colonial rule and other imaginaries of communities have shaped the sense of nationalism in different Bengali communities and how that manifests through their online decolonial discourse.

\section{Literature Review}\label{sec:literature_review}
\subsection{Nationalism as a Collective Identity}
Identity refers to how we see ourselves and want others to see us as social and physical beings~\cite{erikson1968identity, gecas1982self, goffman1959presentation}. While identity is often understood as an individuated construct, this paper focuses on collective identities: identities shaped by people's perceived and actual membership in different social groups~\cite{snow2001collective}. Identity mediates everyday human interaction based on distinguishing characteristics, beliefs, and experiences of individuals or groups~\cite{tajfel1974social}. When connected to broader social and cultural logics~\cite{butler2011gender}, collective identity is expressed and experienced through dimensions such as race, ethnicity, religion, gender, sexual orientation, and social class~\cite{mccall2005complexity}. 

One of the most prevalent ways people express and experience their collective identity is through nationalism. While nationalism is colloquially viewed as a political ideology, we draw on Benedict Anderson, whose seminal work exploring its origins~\cite{anderson2006imagined} defines nations as limited communities containing people with the same interests and traits. According to Anderson, nationalism is a cultural system akin to religious beliefs, offering a sense of continuity in a contingent world. Using the concept of ``imagined community", he explains how people living in modern civilizations imagine connections to other citizens despite the impossibility of interacting with everyone in their society. He argued that nations were not the determinate products of given static sociological conditions such as language, race, or religion but rather imagined into existence despite the impossibility of interacting with everyone within a group. For example, he describes how Muslim communities in various geographic locations, even when they are unlikely to interact, share a sense of unity and camaraderie. He emphasized the sense of connectedness in the minds of the individuals. Prioritizing the role of language in fostering a sense of affinity, Anderson described the importance of institutions like the printing press through mass production, circulation, and standardization of printed materials, which he dubbed ``print capitalism" in shaping and spreading shared identities and imagined communities, in turn developing modern nationalism. Drawing on the example of the dilemma of multilingual European empires, he argues that the idea of similarity in appearance, language, and practice is at the heart of nationalism-based collective identity. As communities undergo different experiences and exposure to different cultures and ideas, their collective identity can evolve and change over time. 

\subsection{How Colonialism Has Shaped Nationalism}
One of the often normalized and invisible mechanisms that have shaped and continue to shape people's experiences and perceptions of nationalism is colonialism~\cite{bhabha2012location, das2022collaborative}. While in the contemporary discourse of development and modernization, nationalism has been relegated as ``a matter of ethnic politics," not too long ago, it was seen as ``one of Europe's magnificent gifts to the rest of the world"~\cite{chatterjee1993nation}. Based on how they viewed nations and how people should be grouped, British colonizers proposed and, in some cases, utilized dichotomous schemes to divide various countries (e.g., two-nation theory in India-Pakistan, two-state solution in Israel-Palestine)~\cite{greenberg2005generations}, and did not consider the complex ways in which people see themselves or see themselves in relation to others. 

Such partitions divided peoples into various nation-states based on Western ideals of nationalism. This is best explored by Partha Chatterjee, who, in his influential work ``The Nation and Its Fragments"~\cite{chatterjee1993nation}, criticized the European ideals of nationalism on which Benedict Anderson developed the foundation for imagined community. Here, Chatterjee argues that nationalism was a colonizing force imposed by ``modular" forms of nationalism imposed by Western powers, questioning whether or not colonized peoples had anything ``...left to imagine?". Modularity, in the context of nationalism, refers to the shared characteristics constituting imagined communities. The creative and powerful imaginations of nations in Asia and Africa, instead of being defined on their own terms--or creating modules of similarity as generated through their own sense of what it means to be a ``collective"--were generated based on difference and the ideals of nationalism as imposed by Western powers. Using anticolonial nationalism in Bengal as evidence, Chatterjee demonstrated that while operating within the Western project of modernity, among the bilingual intelligentsia, the ideation of Bengali nationalism through evolving literary practices and cultural expression differentiated itself as ``recognizably Indian." Though this linguistically grounded nationalism was a strong driving force during different stages of the independence movement~\cite{kaviraj2007making}, at the end of British colonization, religion-based nationalism took priority over Bengali nationalism, leading to Bengal's partition.

This is best illustrated by how Western powers constructed nation-states through divide-and-rule policies. Divide-and-rule policies during the colonial period exacerbated religious animosity, resulting in significant communal violence towards the end of that era~\cite{pandey2001remembering} and ultimately culminating in the formation of two nation-states, India and Pakistan, based on the two-nation theory~\cite{devji2013muslim}. Here, Western powers developed nation-states and subsequently modularized imagined communities based on external perceptions of religion mediated by the central premise that ``...the Hindus and Muslims are two separate nations who cannot live together"~\cite{joshi2010contesting}. The colonial approach to perceiving communities based on a monolithic idea of religion overlooking native sociocultural complexities, nuances, interconnections, values, traditions, and history creates myriad internal and external fractures in the communities' self-perception. Hence, in postcolonial nation-states, repairing identity and ``indigenizing the limbus"--the process of reinventing tradition, language, and culture face diverse conceptualizations of indigeneity and nationhood~\cite{subba2006indigenising}. Multiple possibilities emerge for people to imagine communities across different dimensions, such as language (e.g., Bengali), religion (e.g., Hindu-Muslim), and post-partition nation-states (e.g., Indian-Pakistani-Bangladeshi), and negotiations among one's nationhood across various dimensions (e.g., language, religion, country) are often in flux. For example, in 1947, Muslim-majority East Bengal became part of Pakistan based on its shared religious identity. A few decades later, due to West Pakistan's linguistic-cultural imposition on East Bengal, Bengali nationalism started rising rapidly and replaced Pakistani nationalism to differentiate Bengali Muslims from non-Bengali Pakistani Muslims. Eventually, Bengali nationalism led to the formation of the nation-state Bangladesh in 1971~\cite{schuman1972note}. In post-independence Bangladesh, Madan studied ``two faces of Bengali ethnicity": Muslim Bengali and Bengali Muslim~\cite{madan1972two}, based on whichever identity people put forth. Similarly, the decades-long contemporary struggle of non-Bengali and non-Muslim \textit{adivasi}\footnote{Heterogeneous tribal groups and ethnic minorities across the Indian subcontinent.} communities in Bangladesh highlights the nationalism-based tension across ethnolinguistic and religious differences~\cite{karim1998pushed}. Such prioritization between multi-faceted identities is an individuated concept. In a recent study in the context of the online platform Quora, Das and Semaan studying Bengali people~\cite{das2022collaborative} found similar questions (e.g., \textit{``Are you first a Muslim or a Bengali or a Bangladeshi?"} or \textit{``Are you first a Hindu or a Bengali or an Indian?"}) to be major points of negotiation for identity in relation to their colonial past.

\subsection{Institutions in Shaping Hegemonic Nationalist Discourse}\label{sec:lit_rev_institution_nationalism}
In this paper, we are more broadly interested in understanding why and how people engage in decolonial practices, which is a process wherein colonized territories and societies work to reclaim autonomy and independence from colonial powers ~\cite{fanon2004wretched, laenui2000processes}. However, as previously described, colonization has created incredible complexity around people's perceived imagined communities or nationhood in colonized and postcolonial societies, which can impact people's opportunities or willingness to engage in decoloniality. While postcolonial nation-states like India, Pakistan, and Bangladesh mark a continuity in colonial legacies and institutions, they also embody a rupture from the colonial past, adding complexity to the decolonial discourse. This is further exacerbated by how coloniality shapes not only national identities but also sociocultural shifts, geopolitical processes and interests, and economic institutions within nation-states.

While establishing nation-states and independence from colonial rule is often seen as the end goal of decolonization, scholars have criticized this view as a myopic conceptualization of decolonial objectives~\cite{fanon2004wretched, laenui2000processes}. After the departure of colonial rule, oftentimes elites from the previously colonized communities, whom decolonial scholars described as ``colonized intellectuals"~\cite{fanon2004wretched}, ``interlocutors"~\cite{ansari2020design}, ``bilingual intelligentsia"~\cite{chatterjee1993nation}, occupy the helm of postcolonial nation-state's political and administrative structures. These individuals, through their education and exposure to Western thought and adjacency with colonial rulers, while being members of the oppressed group, had absorbed the ideas, values, and cultural norms of the colonizers--what is known as colonial mentality~\cite{fanon2004wretched, guha1997dominance, nunning2015fictions}. Consequently, governance and other sociopolitical institutions in newly established nation-states often continue to perpetuate and reinforce colonial hierarchies (e.g., through forced integration of smaller ethnic minorities and religious majoritarianism)~\cite{ramnath2012decolonizing, patel2007sociological}. Following the formation of a nation-state, reforming its social, political, and economic structures and practices in a way that reflects the culture and values of the diverse national communities should become decolonization's key objective~\cite{fanon2004wretched}.

Chatterjee argues anticolonial nationalism provides a formula for creating its own domain of sovereignty in political struggle within societies shaped by colonial influences, whether under the rule of foreign colonizers or their native interlocutors~\cite{chatterjee1993nation}. He discussed a fundamental feature of anticolonial nationalism in Asia and Africa~\cite{chatterjee1993nation} wherein it divides the social institutions and practices into two domains--the ``outside" material domain of the economy, state-craft, science, and technology, and the ``inner" spiritual domain bearing the ``essential" marks of cultural identity. He emphasizes preserving the distinctiveness of culture that would guide careful consideration and reconfiguration of technological and material advancements of modernity to reflect previously colonially marginalized communities' values.

Yet, technological and material advancements continue to perpetuate anticolonial ideals and systems of power and privilege. Providing a historicist example, Chatterjee writes, ``An entire institutional network of printing presses, publishing houses, newspapers, magazines, and literary societies is created ... outside the purview of the state and European missionaries, through which the new language is given shape. [They] came to think of its own language as belonging to that inner domain of cultural identity, from which the colonial intruder had to be kept out"~\cite{chatterjee1993nation}. Here, he underscores how anticolonial nationalism contributes to the local communities' cultural expression through media and literary institutions. However, in the construction of national identity during the colonial period, postcolonial scholars have critiqued elite groups establishing hegemony--socio-ideological control over norms, values, beliefs, and institutions~\cite{gramsci1999selections}, while disregarding the views of the subalterns--underserved individuals or groups who experience social, political, and economic marginalization within both colonial and postcolonial societies~\cite{spivak2015can, guha1988selected}. Thus, it becomes important to reflect on ``whose imagined communities" are reflected through institutionalized nationalism. In doing so, Bengali historians studied the political struggle between the elite financially sound dominant caste Hindu landlords and the subaltern financially disadvantaged peasants from underprivileged caste Hindus and Muslims~\cite{chatterji2002bengal, chatterjee1993nation}

Under the influence of a hegemonic nationalist agenda, besides the political structures of the postcolonial nation-states, institutions such as media and education systems continue to perpetuate a power hierarchy that leaves mass colonized populations out of shaping the socio-historical narratives. Scholars have found media representations to perpetuate colonial, prejudiced, and biased messaging~\cite{hall1989cultural}. For example, Indian and Pakistani media have been found to create narratives that alienate religious minorities and reify animosity between neighboring countries~\cite{sahoo2020mounting, mansoor2022decolonial}. Similarly, researchers also criticized how the education systems in this region sometimes propagate certain narratives that teach hyper-nationalism, politically charged historical narratives, and hatred toward neighboring countries~\cite{lall2008educate, yaqian2011defining}. To challenge and subvert the hegemonic narratives and discourses shaped by colonial powers or elite cultures, decolonial scholarship, therefore, recommends decolonial discourse and practices--critical engagements and actions to shape public consciousness and societal narratives about continued colonial impacts on present-day societies. Decolonial activists have highlighted approaches to create counter-narratives, challenge colonial ideologies, amplify diverse voices, and promote critical engagement with media and inclusive pedagogical approaches and curriculum reforms that reflect accurate and inclusive histories, foster cultural pride, and promote a diverse understanding of the world~\cite{lomotey2018bois, laenui2000processes, hall1989cultural, sundaresan2020decolonial}. 

\subsection{Reimagining Communities through Decolonial Discourse on Online Platforms}
In the present digitally mediated era, this leads to the question of how digital media shapes these discourses. Decolonial discourse on online platforms is often mediated through and takes the form of critical conversations on historical and contemporary social, geopolitical, and economic issues and foregrounds diverse local cultural identities and perspectives~\cite{das2022collaborative}. Researchers have explored how technology can facilitate and hinder democratic processes, civic participation, and collective action~\cite{flores2018mobilizing, starbird2019disinformation, lindtner2018design, dosono2018identity}. Looking at grassroots US political movements, Goshal et al. explained how these sociotechnical systems exclude people based on racial, gender, and socioeconomic privileges and shape ideological hegemony~\cite{ghoshal2019role, ghoshal2020toward, sharma2017analyzing}. Scrutinizing this notion of hegemony in the South Asian political landscape, a series of previous works in CHI, CSCW, and ICTD has looked at how religion~\cite{panda2020affording, dash2022insights}, caste~\cite{vaghela2021birds, vaghela2022caste}, social capital~\cite{seth2022closed}, and popularity~\cite{kommiya2022voting} factor into political polarization and marginalization. Besides these empirical studies, a significant body of research focused on designing and evaluating systems that empower citizens. Examples of such endeavors include online civic platforms for deliberating democracy~\cite{semaan2015designing, maruyama2017social}, participatory budgeting as a part of open government and transparency~\cite{kim2015factful, kim2016budgetmap}, and digital activism~\cite{lee2013does, vlachokyriakos2014postervote}.

While digital media platforms (e.g., Quora, YouTube) in many ways support marginalized communities' identity work--the process through which people construct, manage, present, and negotiate their identities~\cite{ibarra2010identity}, these platforms' censorship, copyright practices, and politics impede content creators' freedom and further impede marginalized communities' identity work~\cite{gillespie2010politics, fiesler2023chilling, fiesler2019creativity}. Such duality of sociotechnical systems is reflected in CSCW and HCI scholarship as researchers champion and critique the same platforms, respectively, for their support and hindrance in users' identity expression--in the same study~\cite{haimson2015disclosure, kumar2018uber} or separate ones~\cite{das2022collaborative, das2021jol, ammari2014accessing, ammari2015networked}. This paper is most centrally interested in how digital media contributes to people's engagement in decolonial discourse and identity work that serves as a reclamation project of their local and indigenous identities (note Appendix~\ref{app:local_and_indigenous})--a phenomenon that has been dubbed identity decolonization work \cite{das2022collaborative}. Generally, we know from prior work that digital media can provide opportunities through which marginalized communities can push back against heteronormative societal logics by establishing a community to engage in identity work~\cite{dym2019coming} as well as collective sensemaking during times of conflict~\cite{mark2009expanding, al2010blogging}, amongst others. Our research builds on prior work focused on textual communications and social networks on online platforms (e.g., Reddit, Quora), bringing together a conceptual framework drawing on imagined communities and nationalism to explore people's sociopolitical engagement and identity decolonization work through video-based media. 

Researchers have studied video-based platforms of various content lengths, origins, scales, and content types, such as the short-form video-sharing platforms Tiktok and Vine~\cite{simpson2021you, mcroberts2016viewers, barta2021constructing}, hyper-local platforms like Douyin and Kuaishou~\cite{chen2023my, lu2019feel}, and international and generally longer video-hosting platform YouTube~\cite{yarosh2016youthtube, milliken2008user-youtube}, which is viewed through the lenses of CSCW and other adjacent fields like CHI, science and technology studies (STS), and media anthropology as a video-mediated platform that enables users to upload, watch, and interact with a wide array of user-generated and professional content~\cite{cha2007tube}. Scholarships in these fields analyze YouTube's role in media ecosystems, business models, sociopolitical implications, technological infrastructure, and governance structures' impact on content production and distribution~\cite{burgess2018youtube, ma2023multi, lange2007publicly}.

Studies on user-generated video-sharing online communities have highlighted these platforms' effectiveness in enacting their individuated and shared identities and connecting with others with common interests~\cite{yarosh2016youthtube, milliken2008user-canada, askanius2011online}. Askanius and colleagues deeply explored the use of YouTube videos for sociopolitical purposes~\cite{askanius2011online, uldam2013online}, various genres~\cite{askanius2014video}, evolution~\cite{askanius2016online}, and their convergence with mainstream media~\cite{askanius2010mainstreaming}. Focusing on the representation of cultural practices, Milliken and colleagues assess the role and contribution of user-generated online videos in creating online public sphere~\cite{milliken2008user-canada} and examine their potential and limitations~\cite{milliken2008user-sphere}, particularly in relation to regional identity~\cite{milliken2008user-youtube}. Recent studies and media reports have highlighted how ethnic minorities in non-Western contexts use live streaming and video blogging (\textit{vlogging}) to archive, showcase, and promote intangible cultural heritage activities for cultural sustainability~\cite{lu2019feel, chen2023my, dailystar2023short}.

Scholarship in HCI and CSCW has more broadly contributed to our understanding of the relationship between colonialism and technology in myriad ways, starting with formative work in postcolonial computing and decolonial computing~\cite{irani2010postcolonial, ali2016brief, das2022decolonial}. In our work, however, we focus on decoloniality with a ``historicist sensibility" as recommended by Soden and colleagues~\cite{soden2021time}, through which we can better understand the historic and complex entanglement of race, ethnicity, gender, religion, nationality, caste, and socioeconomic status in the context of the Global South and colonially marginalized native communities~\cite{das2023studying} and how the historical incidents, policies, and perceptions of coloniality-shaped nationalism affect contemporary social practices, cultural identities, geopolitical relationships, economic systems, and challenges both at individual and collective levels and how sociotechnical systems reinforce colonial values~\cite{irani2010postcolonial, dourish2012ubicomp, soden2021time}. 

Among some recent social computing studies that have taken this approach, researchers have studied how racial minorities in the US engage in political conversation in relation to colonial structures and build resilience to retain collective memories and revise colonial narratives using Reddit~\cite{dosono2018identity, dosono2020decolonizing}. Particularly in the Bengali context, Das and colleagues found how sociotechnical platforms (e.g., Quora) reinforce colonial divisions and hierarchies, affecting users' participation and expression of cultural identity~\cite{das2021jol}. In their follow-up work~\cite{das2022collaborative}, Das and Semaan studied how Bengali Quora users collaboratively work toward reclaiming narrative agency for decolonizing their identity. While the decolonial discourse on Quora primarily focused on ethnolinguistic Bengali identities, their work highlighted users' perceived tension across various categorical identities derived from their conceptualized social groups or imagined communities and nationhood, such as the ones defined based on their religion and postcolonial nation-states~\cite{das2022collaborative, das2021jol}.

This paper investigates why and how previously colonized peoples from different nation-states are re-imagining their communities by engaging in decolonial discourse through video-mediated platforms. Here, we draw on Bengali scholar Chatterjee's conceptualization of nationalism~\cite{chatterjee1993nation} and explore how decolonial discourse is shaped by national and collective identities, emerging from the convergence and differences of diverse aspects such as language, religion, and country and how a video-mediated platform like YouTube influences YouTubers' strategies and content.

\section{Methods}
This paper is part of a multi-platform investigation on how computing systems (e.g., Q\&A website, social media, video-sharing platform, algorithmic mechanisms) both support and impede the identity expression of colonially marginalized communities. As colonialism impacted people's identities across the world in myriad and complex ways, challenging the notion of a monolithic local and Indigenous identity and culture, in this paper, we want to understand how people now in different nation-states engage in discourses about their native identity as potentially mediated by and through nationalism. We selected members from Bangladesh, India, and Pakistan due to the multiple waves of foreign subjugation and colonialism that have shaped relationships among people in these countries. This study aims to understand YouTubers' motivations for making videos about colonially marginalized cultures and strategies for video-mediated decolonial discourse. Throughout this exploration, the research examines how these YouTubers envision communities and interpret nationalism, both of which have been influenced by historical experiences of colonialism. To answer these questions, we conducted a qualitative study based on semi-structured interviews with YouTubers who regularly create videos centered around Bengali identity, culture, and people. Before beginning the study, we received approval from our university's institutional review board\footnote{The authors were at Syracuse University during the data collection phase of this study.} for all materials and procedures.

\subsection{Positionality and Reflexivity Statement}\label{sec:positionality}
When researching marginalized communities, the authors' racial and ethnic backgrounds may influence their perspectives and interpretation~\cite{schlesinger2017intersectional, liang2021embracing}. The first author is a cisgender, heterosexual man from the Bengali Hindu minority community in Bangladesh, with a family history affected by refugee crises because of the partition of the Indian subcontinent in 1947 and the liberation war of Bangladesh in 1971. In addition to designing the study, he interviewed participants and led the data analysis. The second author is a cisgender, heterosexual woman from an interfaith minority Jain-Hindu community in India. She helped translate and transcribe several interviews. The third author is an Iraqi-American cisgender heterosexual man from a minority group within Iraq who contributed to the study as the anchor author. He was deeply involved throughout the study from its initial inception and design through the writing of this manuscript. Due to their embeddedness within colonially shaped sociocultural contexts that motivated their research, the authors acknowledge that their work, which focuses on conversations about colonial histories, is inherently political. All authors' memberships in different minority communities, lived experiences in colonially impacted societies, and past experience in critical (e.g., decolonial, postcolonial) computing research motivated their study of colonially marginalized communities' practices around technology. Furthermore, the authors acknowledge that the University of Colorado Boulder, where they are based, is situated on the land of the indigenous Arapaho, Cheyenne, and Ute peoples.

\subsection{Recruitment}
Our study focuses on the video-sharing platform YouTube, which, with over 2 billion monthly users as of early 2023~\cite{globalmediainsight2023YouTubeStatistics}, is one of, if not the most, popular online media in the world. YouTube has a large user following in India (467 million, the platform's largest user base), Pakistan (71.7 million), and Bangladesh (34.40 million)~\cite{statistaYouTubeUsers, datareportalDigital2023}. Its widespread adoption signifies its capacity to capture a diverse range of perspectives and content. It is widely popular among amateur and professional content creators~\cite{yuan2022what}. Moreover, the platform has become a significant space for political discussions, especially among marginalized voices, activists, and individuals passionate about national and global issues. Because of its increasing popularity and capacity to provide avenues for sharing diverse communities' ideas and engaging in discourse, we chose YouTube as the site for recruiting video creators.

Data for this study draws on semi-structured interviews with 15 content creators on YouTube living in Bangladesh, India, or Pakistan. Our eligibility criteria included: participants must be (1) 18 years or older, (2) a creator of YouTube content focused on Bangladesh, India, and/or Pakistan, and (3) had to be living in one of Bangladesh, India, and/or Pakistan. It is important to note that the first author is part of a minority group from Bangladesh, as well as an avid YouTube user. 

We identified participants for our study through a combination of purposive sampling~\cite{suri2011purposeful} and snowball sampling~\cite{goodman1961snowball}. Specifically, prior to recruiting participants, we spent time identifying potential participants whom we could subsequently contact to enroll in the study. Broadly conceived, our purposive sampling strategy included a combination of identifying and recruiting participants by (1) searching for content creators on YouTube and (2) recruiting from personal social networks. Prior works have highlighted that different linguistic, religious, and national identities are central to online decolonial discourse within the Bengali geocultural context~\cite{das2021jol, das2022collaborative}. We searched for YouTube videos using combinations of Bengali identity-related keywords, such as Bengali, Bangladesh, Bangladeshi, India, Indian, Pakistan, and Pakistani, as previously used by~\cite{das2021jol}. Though among the retrieved results, a large number of videos were Bengali movies, dramas, and music, since the focus of this study is discourse, we identified videos structured as dialogues and commentaries. Particularly, we identified channels operated from the region that frequently published videos about geographic and cultural identities. To ensure a more diverse sample, we also relied on YouTube's algorithmic recommendations for related videos and channels. This approach is adapted from prior work~\cite{das2022note}, allowing us to expand the diversity and scope of channels that were part of our larger sample. Moreover, we recruited from personal social networks. After identifying YouTube channels that regularly publish videos about Bengali culture, identity, and people, we considered the YouTubers of those channels as our potential participants.

To enroll participants in our study, we engaged in a multi-site endeavor. Due to the lack of a direct messaging feature within YouTube, we had to engage in myriad activities to contact and recruit participants. Whereas some YouTubers provided contact information in the description section of their channel, others did not. We focused on those YouTube channels where the creators provided a means for contacting them, such as through email or social media. For those channels that did not provide this information, we did not pursue them as it was a challenge to identify alternatives for communicating with them. We collected YouTubers' contact information from those descriptions, which included one or more of the following: email and Instagram, Facebook, and Twitter handle. First, we tried to contact them through email, which included a recruitment flyer describing the study's objectives, participants' eligibility, general information about the interview procedure, and researchers' contact information. Given people's varied email response behavior (e.g., rarely checking for emails), if we did not get a reply from a YouTuber within a week, we sent them a reminder email. If we could not contact a YouTuber through emails, we contacted them on social media from the first author's profiles on corresponding platforms. We also contacted a few potential participants through real-life social networks. Following interviews, we engaged in snowball sampling, wherein we asked participants if they could recommend and connect us with other potential participants. Recruiting and interviewing participants continued until we met theoretical saturation. We enrolled 15 participants in total. We summarize their demographic information like gender, country of nationality, age, religion, education, and occupation in Table~\ref{tab:demography}.

\begin{table}[!ht]
    \centering
    \caption{Demographic information of the participants}
    \label{tab:demography}
    \begin{tabular}{p{1.5cm}p{1cm}p{1.8cm}p{1cm}p{2.15cm}p{1.8cm}p{2.3cm}}
    \toprule
        \textbf{Identifier} & \textbf{Gender} & \textbf{Country of nationality} & \textbf{Age} & \textbf{Religion} & \textbf{Education} & \textbf{Occupation} \\
    \midrule
        P1 & Male & India & 20-24 & Muslim & Bachelor's & Engineer \\
        P2 & Male & India & 35-40 & Hindu & Master's & Journalist \\
        P3 & Male & Pakistan & 25-30 & Muslim & Bachelor's & Student \\
        P4 & Male & Bangladesh & 25-30 & Muslim & Master's & Journalist \\
        P5 & Male & Bangladesh & 30-34 & Muslim & High school & Freelancer \\
        P6 & Female & India & 30-34 & \textit{Did not disclose} & Bachelor's & TV presenter \\
        P7 & Male & Bangladesh & 30-34 & Muslim & Bachelor's & YouTuber \\
        P8 & Male & Bangladesh & 30-34 & Muslim & Master's & Journalist \\
        P9 & Male & India & 40-44 & Hindu & Master's & Govt. employee\\
        P10 & Male & Pakistan & 25-30 & Muslim & Bachelor's & Engineer \\
        P11 & Female & Bangladesh & 25-30 & Muslim & Master's & Job Aspirant \\
        P12 & Female & India & 20-24 & Hindu & Bachelor's & Student \\
        P13 & Male & Pakistan & 30-34 & Muslim & Master's & Engineer \\
        P14 & Male & India & 20-24 & Hindu & Master's & Web-developer\\
        P15 & Female & India & 20-24 & Hindu & Master's & Student\\
    \bottomrule
    \end{tabular}
\end{table}

\subsection{Interviews}
Following the qualitative methodology outlined by Strauss and Corbin~\cite{strauss1994grounded} and Yin~\cite{yin2017case}, we conducted 15 in-depth semi-structured interviews between the Summer of 2020 and the Summer of 2022. Interviews lasted from 30 minutes to 1 hour and 53 minutes (averaging approximately 60 minutes). Given that all of our participants were international and physically distributed across different regions of Bangladesh, India, and Pakistan, interviews were conducted using the technology that was most comfortable and accessible to participants, including Zoom and telephone. In some cases, when the Internet connection was unstable, we switched from Zoom to telephone. Participation in this study was voluntary, meaning participants did not receive remuneration for their participation. Since we initially intended for the interviews to last only one hour, we alerted participants when we reached that time limit. However, all of them wanted to continue beyond the pre-decided duration. Prior to initiating any given interview, we read an oral consent form to the participants. All participants provided verbal consent and also consented to the interviews being recorded. Importantly, the first author is a native Bengali speaker with bilingual proficiency in English and a working oral proficiency in Hindi/Urdu\footnote{As spoken languages, Hindi and Urdu are mutually intelligible, to the point that they are sometimes considered dialects or registers of a single spoken language known as Hindi-Urdu or Hindustani~\cite{editors2022hindustani, wiki2022hindi}.}. He conducted all the interviews in Bengali, English, or Hindi/Urdu based on participant preferences. 

Interviews were designed as life histories~\cite{wengraf2001qualitative}, seeking to understand and locate people's video content creation within their long-term life experiences. We initiated interviews with demographic questions, followed by questions seeking to understand how their family and life histories have shaped their perception of their sociohistorical perspectives. We then asked participants about their motivations and online and offline experiences for making videos on Bengali culture, history, and society, followed by questions about their video-making practices. Each interview, on average, was approximately an hour long. We transcribed the interview recordings and translated the non-English interviews into English. The first author translated the Bengali interviews, while the second author, a native Hindi speaker, did the same for the Hindi/Urdu interviews. All transcripts were anonymized and de-identified before analysis.

\subsection{Data Analysis}
We analyzed our data using an inductive, grounded theory~\cite{strauss1994grounded} inspired approach, commonly used in qualitative HCI and social computing studies~\cite{tushar2020we, houston2016values, das2021jol}. We used qualitative data analysis software MaxQDA Plus to code the interview transcripts. Following Strauss and Corbin's guidelines~\cite{strauss1994grounded}, we analyzed our data in three phases. In the open coding phase, the first author iteratively reviewed the interview transcripts and identified the repeatedly appearing abstract concepts, events, and interactions. Some examples of open codes emerging in this phase are: \textit{``ancestral ties and cultural connection"}, \textit{``childhood exposure to communal violence and political turmoil"}, and \textit{``living memories of partition and migration as refugees"}. We associated quotes from the participants with the corresponding open codes. The first author conducted the open coding and continued to meet with other members of the research team on a weekly basis to discuss the emergent codes. We then collaboratively engaged in axial coding, where we combined the open codes to create higher conceptual themes. For example, aforementioned open codes were merged to create the axial code \textit{``personal ties and lived experiences"}. Finally, we identified the relationships between the axial codes in the selective coding phase, resulting in the themes, i.e., motivations and strategies, we present in this paper. Since interview data is contextual, our reflexive and interpretivist grounded theory-based data analysis~\cite{strauss1994grounded, sebastian2019distinguishing} does not require us to calculate an inter-coder reliability score~\cite{mcdonald2019reliability}. Moreover, it is important to note that while our work stayed true to interpretivist tradition, that means the analysis was not only ``about" the phenomenon under study but equally, implicitly or explicitly, ``about" the perspective of the researchers~\cite{anderson1994representations}. Thus, in addition to the context of the work and its focus on conversations about colonial histories, which is inherently political, the authors' unique demographic and scholarly backgrounds (discussed in section~\ref{sec:positionality}) have shaped the decolonial interpretations of the data that are presented in this work.

\subsection{Limitations, Future Work, and Ethical Considerations}
Postcolonial computing scholars have argued that sociotechnical systems such as YouTube being designed in the West can impede the cultural and identity expression of users in the Global South~\cite{irani2010postcolonial, das2021jol}. While this paper primarily examines YouTubers' motivations and strategies for decolonial discourse in the context of Bengali culture, we did not look at the algorithmic side of the platforms where these discourses take place. Sociotechnical systems (e.g., recommendation algorithms) can present unique politics by perpetuating algorithmic coloniality or hyper-nationalism~\cite{das2021jol}. In our future work, we will study the challenges users face using YouTube and the ways in which they navigate through those issues. Another limitation of our study is the lack of gender diversity among participants. Similar to Das and colleagues' paper on Bangladeshi sociopolitical influencers on social media~\cite{das2022understanding}, we found fewer female YouTubers making videos in the space of sociopolitical decolonial discourse, possibly due to the tense political environment in the region~\cite{jalal2009democracy}. We also found only a few content creators who came from religious minority backgrounds. In addition to videos on local culture being a niche topic, recent work found that Bangladeshi minority communities' experience on social media is shaped by fear and a spiral of silence~\cite{rifat2024politics}. Despite our several strategic attempts, the nature of the content creation space constrained us from recruiting a Hindu participant from Bangladesh while interviewing only one Muslim participant from India. Therefore, the first author, who identifies as a Bangladeshi Hindu, was especially mindful of this group during the study, given how the religious majority-minority composition often relates to political power in South Asian nation-states formed based on the two-nation theory. Moreover, due to the conservative subcontinental culture, despite our several focused attempts to recruit more female participants, we could interview four female YouTubers. Among them, two participants requested to be interviewed together with their YouTube channels' co-patrons in the same call (one cited following Muslim guidelines for women socializing with non-familial men as the rationale for her request). While interviews in a group setting created a possibility for some participants to suppress their opinions or for one participant to dominate the conversation, we did not find any visible hesitancy. The interviewer also strategically navigated the conversation so that all participants in those calls were equally responsive. Again, all but one of our participants had at least graduated from university, leading to the possibility of the study reflecting views of higher educated people over the mass population. While all our participants belong to previously colonized communities, they are from age groups that have not experienced British or Pakistani colonial rule themselves. Therefore, future work should look into understanding the experiences of people who not only belong to previously colonized communities but also experienced colonial subjugation themselves or faced colonially created crises (e.g., being refugees due to partition or war). Colonization, when viewed as a crisis~\cite{das2022collaborative}, researchers should also consider the possible risk of older participants reliving traumatic experiences of the colonial past. Since our participants drew on experiences of living in colonially marginalized communities and the ones they heard from older family members, their risk of reliving such traumatic experiences was minimal. Moreover, our recruitment of participants is heavily influenced by our search for relevant videos and channels on YouTube. Similar to most qualitative research~\cite{leung2015validity}, this paper aims not to produce generalizability but rather to study a specific process in a defined context.

\section{Results: Re-Imagining Communities through Decolonial Discourse}
During our interviews, our participants openly shared their family history and personal lived experiences. Throughout these conversations, we discovered that the partition of Bengal into different nation-states led to isolation and communication gaps among the Bengali people. As a result, our informants described ``using their imagination" to develop perceptions of their imagined Bengali community. Our participants discussed how their everyday experiences motivated them to delve deeper into their culture and history. For example, Participant P12 explained her perception of transnational collective Bengali identity, driven by her family history and ancestral ties:

\begin{quote}\textit{
    We often talk about Bangladesh in our house. Ancestors of both my parents are from Bangladesh. My father grew up in [a Bengali-speaking Indian state]. There is a place called X where he lived with my grandmother and his five other siblings. They struggled a lot after migrating from Bangladesh to India. My mother grew up in [another Bengali-speaking Indian state]. Because the Bengali people have a very strong attachment to Bangladesh, I am interested in Bangladesh.
}\hfill (P12, female, India)\end{quote}

Participants discussed how their multi-faceted linguistic dialects and internalized cultural customs strengthened their connection and attachment to people in neighboring countries beyond physical borders. This connection to their Bengali identities and communities that were dislocated due to historical colonialism has led to two broad forms of decolonial discourse on YouTube that fall in line with Chatterjee's assertion that nationalism drives anticolonial practices in both the ``inner" domain of cultural identity and the ``outside" domain of institutionalization~\cite{chatterjee1993nation}. We explore the motivations of YouTubers' video-mediated decolonial discourses as characterized by their experiences with (1)~nationalism's impact on the cultural community, the ``inner" domain of cultural identity, and (2)~the ``outside" domain of nationalism defined through institutionalization, where their strategies include making different kinds of videos, such as travel vlogs, social interviews, reaction and journalistic videos, political explainers, and satires. We also elaborate on how each kind of video contributes to the Bengali decolonial discourse on YouTube.

\subsection{Nationalism's Impact on Cultural Community: The ``Inner" Domain of Cultural Identity} 
Conceptualizing and committing to a vision of native culture and national identity are vital to decolonization~\cite{laenui2000processes, fanon2004wretched}. We examined YouTubers' reflections on how the colonial history of Bengal has socio-psychologically conditioned people in this region to perceive the ``inner" domain of cultural identity--their collective native identity--and how they strategize their video-making to bridge isolated and disintegrated conceptualizations of Bengali culture and identity.

In the sections that follow, we first describe people's motivations for producing YouTube discourses centered around their collective identities. We then describe the decolonial strategies they employ as part of their identity decolonization work.

\subsubsection{Motivations}
Our findings highlight how participants use language-based culture, religion, and post-partition nation-states as threads for identifying fractures. By negotiating collective identities, they foster affinity beyond national borders. We elaborate on how their transnational conceptualization of communities and interaction with their viewers affect their content creation and engagement. We present their motivations through the following themes: (a) language-based cultural similarities and diversity, (b) religion driving reconciliation and empathy, and (c) negotiating identities based on postcolonial partitions.

\noindent\paragraph{\textbf{Re-Imagining Community Around Language-based Cultural Similarities and Diversity}}\label{sec:language}
Bengali YouTubers highlighted celebrating the similarities and diverse ways in which people express and practice local ethnic identities and associated linguistic and cultural practices in Bangladesh and India as their motivation behind making videos on YouTube about their culture. 

Despite their community being displaced across the fifth longest and one of the most dangerous international borders in the world~\cite{tbs2020bangladesh}, the YouTubers in these two countries were motivated by the opportunities to re-imagine their community based on their language; they wished to celebrate and highlight their linguistic identity as a common feature of their displaced community. Participant P15 described her rationale and strategy to foreground her Bengali identity through her YouTube channel:

\begin{quote}\textit{
    We are Bengali, and we speak in Bengali. Since the national language of Bangladesh is Bengali, they [Bangladeshis] would be able to connect with and relate to our videos. ... We always try to prioritize the Bengali language in our channel. So, we named our channel X [explains the name's etymological relation to Bengali]
}\hfill(P15, female, India)\end{quote}

The associated cultures and customs as shaped through linguistic norms also inspire the YouTubers to connect with people in \textit{``Opar Bangla"}\footnote{Given the colonial history of partition of Bengal, Bengalis in Bangladesh (then East Bengal) and Indian state of West Bengal refer to their side of Bengal as \textit{``epar Bangla"} (``this side of Bengal") and the part of Bengal on the other side of the Bangladesh-India border as \textit{``opar Bangla"} (``the other side of Bengal"). These phrases are usually used in cultural and literary discussions~\cite{chattopadhyay2022role, banerjee2015more}.}. These YouTubers imagine themselves and their audiences as part of a broader community of Bengali identity around the linguistic dimension and beyond geographic boundaries. Examples include Bengali norms of addressing each other even when they have never met as \textit{``dada"} (brother) or \textit{``didi"} (sister) that foster closeness with their audiences. Participant P9 described how his perspective drives his intent to be on YouTube:

\begin{quote}\textit{
    We became two different countries, two different peoples. In fact, I did not know much about them [Bangladeshi Bengali audiences]. I became closer and more cordial with them when I got to know them better. I was fascinated by their manner and my conversations with them. ... The space of cordiality widened among us.
}\hfill(P9, male, India)\end{quote}

The diversity of dialects introduced by several migrations between two sides of Bengal creates a probability for some Indian Bengali (e.g., P12) and some Bangladeshi Bengali (e.g., P7) YouTubers talked about having similar dialects as Bangladeshi Bengali and Indian Bengali audiences, respectively. Those participants think that this builds a perception of belonging to a further closer group based on similar dialects (e.g., \textit{Bangal} and \textit{Ghoti}\footnote{Bangal and Ghoti are distinguishable variations of the Bengali language spoken by people generally in the Eastern and Western region, respectively~\cite{das2021jol}.}), within the broader language-based imagined community, motivating them to create content on diverse regional linguistic and folk cultures.

\noindent\paragraph{\textbf{Towards Religious Reconciliation and Empathy: Re-Imagining the Role of Religion in Community}}\label{sec:religion}
Our participants acknowledged how religion had shaped their sense of community through colonization. They were motivated by the opportunity to reimagine the role of religion from one that created division to one that allows for reconciliation and empathy-building.

All our participants from Pakistan (P3, P10, P13) and some from Bangladesh (P4, P5) who belong to the Islamic faith have highlighted their religion as a strong aspect of their identity and see that as a connecting thread among people in transnational contexts beyond borders. To describe how his religious value motivates him to cooperate with other Muslims, P13 commented,

\begin{quote}\textit{
    As Muslims, we need to help each other. As our Prophet (PBUH) said that a Muslim is a brother of another Muslim. ... The question always comes up ``What did you Muslims do in 1971?" ... There are always clashes between brothers on some issues. But that doesn't mean you should get separated from each other and never connect.
}\hfill(P13, male, Pakistan)\end{quote}

The quote from this Pakistani YouTuber essentially uses religion as the basis for defining communities. He also promptly responded to a common critique of religion-based nationalism in the context of Pakistan's oppression of Bengalis despite common religious identity and the genocide of Bengalis in 1971. We found similar sentiments among other Pakistani participants who make videos about Bengali culture. Their perspectives echo prior scholarships on Muslim identity and how it differentiates between believers (Muslims) and non-believers (non-Muslims) as a gatekeeping mechanism for imagining communities~\cite{ambedkar1979dr, anderson2006imagined}. Based on the Muslim Brotherhood, a strong willingness for reconciliation between Bangladesh and Pakistan was visible in their conversations. The same focus on harmony and cooperation is also present in these YouTubers' videos, especially in cases of easier visa processing and increased financial cooperation.

Some Bangladeshi and Pakistani YouTubers (P4, P11, P13) highlighted how their common belief in Islam establishes a sense of community with Muslims within Indian borders. For example, Participant P4 shared his experience of visiting Kashmir, which is one of the Muslim-majority Indian union territories~\cite{statistaIndiaMuslim}, and interacting with people there:

\begin{quote}\textit{
    When [Kashmiri people at the hotel] heard that we went there from Bangladesh and we were Muslims, they were saying that they were also Muslims, and so, we were brothers. They are hospitable and sweet talkers.
}\hfill (P4, male, Bangladesh)\end{quote}

Similarly, some Indian YouTubers (P2, P12) expressed their sense of affinity with Bangladeshi Hindus based on their religious identity. In these cases, the religious similarities made the YouTubers compassionate and empathize with the struggles of people of the same religious identities in different countries and motivated the YouTubers to focus on religious minorities like Bangladeshi Hindus and Indian Muslims, for example, resulting in Bangladeshi and Pakistani YouTubers talking about Indian Muslims' marginalization and Indian YouTubers sympathizing with Bangladeshi Hindus during the times of religious violence. In some cases, this translated to empathizing with minorities in their own countries. For example, P2, an Indian Hindu YouTuber, described his effort to highlight the injustice to an Indian Muslim woman despite many likely being indifferent to it:

\begin{quote}\textit{
    We will release an episode on Bilkis Banu today. I know it won't do even half a million [views] because who wants to know about the 2002 Gujarat riots and the rape of a Muslim woman there.
}\hfill(P2, male, India)\end{quote}

As religion shapes participants' imagined communities, their videos bring forth minorities' experience with regional geopolitics (e.g., Kashmir), the rise of religious extremism, violence, and minority persecution--transgenerational issues tracing back to colonial divide-and-rule policies.

\noindent\paragraph{\textbf{Re-Imagining the Postcolonial Partition: Towards Understanding Geopolitical Neighbors}}
In contrast to the participants we mentioned so far, whose sense of imagined community motivates their engagement in decolonial discourse, some YouTubers were motivated to produce videos that continued to perpetuate the colonial logic of partition that divided the Bengali people. They persist in the nationalist agenda that continues to impose a Western modularization of nationalism based on their specific nation-states. For example, Bangladeshi YouTubers often motivate their video-making practices from the point of Bangladeshi nationalism. Participant P4 described how Bangladeshi nationalism served to motivate his video production on YouTube:

\begin{quote}\textit{
    Bangladeshis watch lots of videos about what foreigners know or think about Bangladesh. This makes it clear that nationalism is strong among Bangladeshis. If someone praises Bangladesh, they become happy and listen to them attentively.
}\hfill(P4, male, Bangladesh)\end{quote}

Similarly, a minority of Pakistani (P10, P13) and Indian (P14, P15) YouTubers described an impulse to propagate Pakistani and Indian nationalism. These YouTubers were deeply motivated by opportunities to imagine exclusive communities defined based on modern nation-states. These YouTube channels were shaped by hyper-nationalist views that harbor collective hatred towards individuals from other countries. Participant P4, the same individual as above, explained his perspective on post-independence Bangladeshi nationalism:

\begin{quote}\textit{
    Because of our history of the liberation war, whenever we talk to a Pakistani, the response your brain automatically gives you is that they are our enemy, they oppressed us, they dishonored our mothers and sisters, and they killed our fathers and brothers. This reality will shake you at the very beginning. 
}\hfill(P4, male, Bangladesh)\end{quote}

Here, the Bangladeshi Participant P4, who empathized with Kashmiri Muslims in India based on religion, is prioritizing his nation-state-based Bangladeshi identity over religious identity, focusing on the country's history of the liberation war against Pakistan. This exercise of nationalism, as Chatterjee argued~\cite{chatterjee1993nation}, leads the YouTubers to imagine communities based on differences. As an example of that, Bangladeshi nationalism is often defined through actively distancing themselves from Pakistani identity. In contrast, the postcolonial partition of the subcontinent in 1947, where Bangladesh used to be East Pakistan, influenced the Pakistani YouTubers' (P10, P13) imagination of a community. Participant P10 described his view:

\begin{quote}\textit{
    Whenever I hear the word Bangladesh, the first thought that comes to my mind is that it was a small part of us, known as East Pakistan, which later got separated. So, thinking about it now, it would be great if we were still together.
}\hfill(P10, male, Pakistan)\end{quote}

From the angle of Pakistani nationalism, this participant mourns about Pakistan's history with Bangladesh, which Laenui defines as an important step in the decolonial process~\cite{laenui2000processes}. Echoing South Asian scholars' argument that Pakistani nationalism builds upon collective Muslim identity that separates itself from other religious communities~\cite{ambedkar1979dr, devji2013muslim}, the Pakistani YouTubers strongly tied their sense of nationalism to their strong remembrance of the two-nation theory. Contemporary political relations, interests, and conflicts and the rise of far-right exclusionary nationalism like Hindutva in India~\cite{nandy1989intimate} and Islamism in Bangladesh~\cite{riaz2004god}, which are distinct from anti-colonial nationalism, also influence people's perception. Besides making YouTube empathize with minorities in other countries, as we previously discussed in section~\ref{sec:religion}, some participants' use of negative phrases like ``enemy" and ``conspirator" to describe neighboring countries demonstrated how those factors accentuate the process of othering.

However, the majority of our participants highlighted how their motivations were guided by anticolonial logic and a decolonial impulse to re-imagine their post-partition relationships with their neighbors by seeking to understand their geopolitical neighbors' knowledge and perspectives about them. For example, participant P11 evaluated her audience's curiosity about other nation-states as an opportunity to develop mutual cultural and historical understanding:

\begin{quote}\textit{
    Especially when an Indian or a Pakistani person talks about Bangladesh’s history and culture or what they know, for example, [about] what happened during the liberation war, or about our linguistic tradition or views on how Bangladesh is developing, people become interested.
}\hfill(P11, female, Bangladesh)\end{quote}

Other YouTubers were motivated to portray their countries in a more positive light and explain their perspectives to their geopolitical neighbors. Given the negative attitudes from neighboring countries like Bangladesh and India toward Pakistan, some Pakistani YouTubers use the platform to advocate for their country. Participant P13 to describe his motivations, said,

\begin{quote}\textit{
    Every person is an ambassador of his country. ... Even South Asian countries ... have negative thoughts about Pakistan. And we were wondering why so. We are the people that others have negative thoughts about. So, we thought, ... we would promote our culture, our values, and a positive image of our country.
}\hfill(P13, male, Pakistan)\end{quote}

Some of our participants (e.g., P11, as quoted earlier) believed that promoting their countries to a broader regional and international audience serves to manage how others see or feel about their collective identity, which makes their videos more likely to be popular among audiences from their countries. This pragmatic potential to gain popularity motivates YouTubers to make videos about their national identity and local cultures. While the decolonial discourse and practices should embody the ``essential" marks of cultural identity due to colonial history and partitions, the Bengali culture and collective identity have been fragmented in the imaginaries of the people in this region.

\subsubsection{Decolonial Strategies and Re-Imagining the ``Inner" Domain of Cultural Identity}\label{sec:decolonialstrategies1}
To realize their stated motivations centered around revising existing linguistic, religious, and partition-based collective understandings, YouTubers adopt decolonial strategies to re-imagine the ``inner" domain of cultural identity. Particularly, YouTubers were working to re-imagine and establish a new collective Bengali identity on their own terms. The strategies employed include (a) fostering transnational cultural understanding through reaction videos and social interviews and (b) decolonial storytelling using travel video blogs at historical sites and cultural festivals as probes.

\noindent\paragraph{\textbf{Reaction Videos and Social Interviews to Foster Interconnected Cultural Understanding}}\label{sec:reaction_videos_and_social_interviews}
Our participants often make videos about cultural artifacts and practices of their neighboring countries to foster an interconnected regional cultural identity, which are often structured as reaction videos, social interviews, or public reactions. They make videos showing their emotional reactions while watching entertainment artifacts like television series episodes, film trailers, music videos, and short documentaries. These videos are colloquially called ``reaction videos" and are numerous and popular on video hosting services such as YouTube~\cite{wikipedia2020reaction}. Through these videos, our participants try to identify and highlight intercultural similarities among Bengali communities' cultures in different countries. As discussed earlier, they often focus on language, religion, or postcolonial partitions as axes while finding similarities. For example, participant P3 said:

\begin{quote}\textit{
    My videos that became popular were mostly videos about the Islamic scholar [name]. He is a famous scholar of Bangladesh. I get a lot of views on those videos. Many Pakistani and Bangladeshi people like those videos.
}\hfill(P3, male, Pakistan)\end{quote}

Here, the participant described his strategy to understand his Bangladeshi viewers' cultural practices using religion as a lens. Given the cruciality of religion in the region, our participants talked about accommodating different religions' various influences within their culture instead of the ``thoughtless adoption of European customs". To do so, they made videos on Muslim festivals Eid-ul-Fitr and Eid-ul-Adha, Hindu festivals Diwali and Holi, and common Bengali festivals like \textit{``Boshonto boron"} (welcoming the Spring season). In other videos, some YouTubers (P13, P14, P15) have also made such reaction videos about historical figures and events (e.g., the language movement of 1952), popular personalities, cultural topics, and artifacts. Since assessing the potential popularity of a topic in a different country can be difficult, our participants check out trending topics before making videos. Participant P1 explained his process as follows:
 
\begin{quote}\textit{
    I will go to the internet, and first, I will search for topics that are trending and suitable for my channel. I am now preparing a video about Bangladesh, and I am writing the script. A week ago, I searched on Google for ``trending topics in Bangladesh."
}\hfill(P1, male, India)\end{quote}

Another type of video several of our participants make to foster a broader understanding of Bengali culture among people from transnational contexts is structured as social interviews. In these videos, YouTubers go to public places and ask random people about their knowledge and perceptions of people in neighboring countries. Our participants used phrases like ``social experiment" or ``public reaction" to describe this genre of videos. Participant P8 described the questions he asked while making some past ``public reaction" videos:

\begin{quote}\textit{
    When we made videos in Kolkata or India, we asked, for example, what the people in Kolkata know about Bangladesh. Who are real Bengalis--the people in Bangladesh or the people in West Bengal [, India]? Or what do the people in Pakistan know about Bangladesh cricket?
}\hfill(P8, male, Bangladesh)\end{quote}

To assess how much the people in their countries know about their neighboring countries, our participants use various physical artifacts (e.g., other countries' flags and currencies). Participant P10, shared his strategies for ``social experiments":

\begin{quote}\textit{
    I made basic videos, such as videos on Bangladesh's currency or flag. I would show some countries' flags to people and ask them which was Bangladesh's flag. ... I gradually went deeper and made more insightful videos about Bangladesh.
}\hfill(P10, male, Pakistan)\end{quote}

Like reaction videos on entertainment artifacts, our participants emphasize finding similarities through these social interviews. Interestingly, the sport cricket as a postcolonial influence~\cite{majumdar2007nationalist} excites people and serves as a connecting point for Bangladesh, India, and Pakistan. Their common food culture also brings people from these countries together. Participant P15 mentioned a recent video, in which she focused on the similar food cultures and ambiances in Bangladesh and India:

\begin{quote}\textit{
    We are making another video today on Bangladeshi street foods. We make videos about Dhaka. We shot a video yesterday at Purbachal\footnote{Among others, this place has recently been hyped for many food options that are good for groups, have all-you-can-eat options, or are warm and cozy.}. That place seemed similar to our new town in Kolkata.
}\hfill(P15, female, India)\end{quote}

Here, the participant describes how she perceived two neighborhoods, one Bangladeshi and another Indian, as being similar based on comparable availability and popularity of street food. Gathering at street corners over tea or street food to discuss intellectual matters or gossip among intimate friends, which postcolonial scholar Dipesh Chakrabarty dubbed as \textit{``adda"}~\cite{chakrabarty2008provincializing}, in his words, is ``something quintessentially Bengali, ... an indispensable part of the Bengali character"~\cite{chakrabarty2008provincializing}. Similar to prior work~\cite{mark2009expanding} that found videos highlighting people's everyday practices to be helpful in humanizing people in different social contexts, through these videos highlighting cultural practices and artifacts, our participants aim to facilitate a better cultural understanding by disproving myths and stereotypes about Bengali communities across borders and religions, shaping a more harmonious attitude toward each other.

\noindent\paragraph{\textbf{Decolonial Storytelling through Travel Vlogs to Highlight Local History and Culture}}
While recent scholarship in CSCW and CHI has discussed the role of video blogs (\textit{vlogs}) and live streamings in cultural sustainability and representation of intangible cultural heritages, rituals, and practices~\cite{chen2023my, lu2019feel}, our participants prioritize historical perspectives besides cultural festivals in their vlogs. Those who make vlogs (P9, P14, P15) often choose to structure those as travel vlogs--a particular type of vlog where they showcase and self-document their journeys to places of Bengali historical importance. For example, participant P9 described one of his popular vlogs:

\begin{quote}\textit{
    Initially, I tried to present my videos like travel vlogs where I would show a new place to people on YouTube. In the early days of my channel, in a video, I showed indigo trees. I read a lot about the indigo plantations in Bengal by the British. But how many of us even recognize indigo trees? General people do not even know that there are a lot of indigo trees around us. I used that indigo tree as a ``tool" to make one of the initial videos about British colonial rule--how they oppressed and forced the farmers to cultivate indigo. I got over 1.5 million views on that video.
}\hfill(P9, male, India)\end{quote}

Here, the participant explained his attempt to highlight the Indigo plantation, which was a significant factor in the historic devastation of both East and West Bengal's agriculture, yet is a globally lesser-highlighted aspect of British colonization in the Indian subcontinent. In doing so, he used an everyday material probe--an object inspiring open-ended and evocative activities like story-telling~\cite{wyche2020using}. Such probing invites viewers to revitalize their historical sense and memory of the British colonial period, which is a critical component of decolonization discourse online~\cite{das2022collaborative}. As the history of common glory and, more importantly, that of common sufferings brings a nation together~\cite{renan2018nation}, these vlogs can evoke memories of past glories and sufferings among Bengali people, now living in different nation-states, leading them toward bolstered and interconnected Bengali identity. Many of our participants were born and raised in historic towns and villages--a fact in which our participants took pride. They often make these travel vlogs in sites neighboring their living places, with numerous instances of traveling further, such as major tourist attractions. Their focus on local historical sites underscores the deep-rooted nature of colonial influences, makes it relatable in people's everyday experience, and contributes to constructing historical narratives from local points of view. In Participant P9's words:

\begin{quote}\textit{
    Through YouTube, I can show all the nearby places and explain their history to people. I started from the often comparatively ignored places and their history.
}\hfill(P9, male, India)\end{quote}

These local perspectives complement people's historical understanding and add nuances to the native narratives of colonization in an attempt to foster a new and re-imagined Bengali collective identity. To surface a sense of historical reminiscence, our participants (e.g., P9, P14) make travel vlogs at old temples and mosques. As an alternate strategy to feature such Bengali cultural festivals, some of our participants also make travel vlogs at various iconic places (e.g., Rabindra Bharati University\footnote{Rabindranath Tagore is a Noble laureate and famous Bengali poet. Rabindra Bharati University is a public university located at his family home, Jorasanko Thakur Bari, in Kolkata, India, founded to commemorate Tagore's birth centenary.}). They believe that these physical explorations, self-documentations, and sharing of experiences serve as points of historical and cultural reflection and realization for themselves and their viewers. As probes (e.g., place, everyday objects) are effective in fostering decolonial storytelling about native and indigenous cultures~\cite{rosner2019whose, charlotte2020decolonizing}, they described YouTube as a more engaging and effective medium than textual blogs and encyclopedias for historical and cultural storytelling.

\subsection{Nationalism Defined through Institutions and Establishments: The ``Outside" Domain of Institutionalization}
Our participants discussed the nationalist agenda perpetuated by different institutions in their postcolonial nation-states. As institutionalized nationalism reinstates colonial values and reflects hegemonic perspectives, they also described their motivations and decolonial strategies, using video creation as a means to resist hierarchies and power imbalances.

\subsubsection{Motivations}
Since various institutions in modern nation-states perpetuate colonial legacies, our participants were motivated by the opportunity to re-imagine the institutions that have shaped their collective imaginaries of what it means to be a community. We elaborate on this through the following motivations, including (1) identifying and resisting colonial impulses within sociopolitical structures, (2) revising education systems' fragmentary and selective historical narratives, and (3) re-imagining transnational representations in mainstream media. YouTubers elaborated how their observations of these institutions falling short of constructing comprehensive historical narratives motivated them to initiate decolonial discourse. 

\noindent\paragraph{\textbf{Identifying and Resisting Colonial Impulses within Sociopolitical Structures}}
Chatterjee argues that through colonization, nationalism in the Indian subcontinent materializes through differentiation~\cite{chatterjee1993nation}. This differentiation wherein nation-states are weaponized against one another is reinforced by existing sociopolitical institutional structures, including politicians, numerous armed conflicts at borders, and adversarial exchanges at international venues are manifestations of nationalism through institutions~\cite{gillan2002refugees, ganguly1996explaining}. To address these transgenerational colonial impacts~\cite{das2022collaborative}, our participants are motivated by opportunities to reimagine regional geopolitics and find ways to highlight the public desire for harmony and cooperation. For example, participant P10 said:

\begin{quote}\textit{
    The more fights there are between the two [countries: India and Pakistan], the more problems it will create for the respective countries. Money will also not be wasted on such wars, which can otherwise be used for better purposes, and people in both countries will be happier. ... for people in Pakistan and India, 70-75\% people are against any kind of violence and want to live in peace. Both countries will only grow when 99-100\% of people think wars should not happen.
}\hfill(P10, male, Pakistan)\end{quote}

Here, he criticized the India-Pakistan relationship and explained how his country's political narrative differs from most people's perspectives. As regional political narratives, contemporary events of communal extremism, intolerance, and the rise of religious majoritarianism carry on colonial legacies of religion-based partition, Participant P2 shared how institutionalized religion in regional politics affected his childhood memories and motivated his discourse acts on YouTube:

\begin{quote}\textit{
    My earliest political recollection was the demolition of Babri Masjid\footnote{A long-standing dispute over a religious site in Ayodhya, India, involving competing claims between Hindus and Muslims regarding the construction of a mosque or a temple at the location.} when I was 12 years old. I remember the curfew that we lived under for two weeks at least. Having seen the debate around it since school. I have always been interested in politics since school, not being a politician, but reporting and talking about the politics of it.
}\hfill(P2, male, India)\end{quote}

Similar to P2, whose childhood experiences motivated him to build political awareness, two of our Pakistani participants talked about the country's internal politics around Bengali people and this diaspora community's nationality and rights being denied.

Besides discussing these colonially designed social injustices, participants (P1, P2, P13) were further motivated by opportunities to examine how the institutionalization of coloniality has continued to shape their everyday experiences and existence, which are crucial decolonial practices~\cite{fanon2004wretched, laenui2000processes}. They purposefully made videos about economic struggles, reformation, and development to underscore the ongoing impacts of coloniality on relationships with their neighbors. For example, participant P1 described his interest in identifying economic impediments:

\begin{quote}\textit{
    Why are we so lagging behind? Why is there so much poverty in our countries? Why are they in so much economic distress? Why is there so much corruption? We have to look into all those and find common ground so that our people can prosper.
}\hfill(P1, male, India)\end{quote}

In doing so, YouTubers identified colonialism's continuing impacts, such as adversarial geopolitics, social injustices, and economic hurdles in subcontinental establishments, which they deem to be the prerequisite to finding ways to remain proactive in addressing those issues. They also scrutinized how colonial values and structures are reinforced through education systems and media.

\noindent\paragraph{\textbf{Revising the Education Systems' Fragmentary and Selective Historical Narratives}}\label{sec:education}
Like Chatterjee highlighted the role of print media and literary works in constructing historical narratives and the idea of nations~\cite{chatterjee1993nation}, we found that most of our participants were motivated to produce YouTube content to revise the narratives being produced by educational institutions. Informants described the role of textbooks and historical narratives featured in their national education system as an important factor in making them interested in Bengali sociohistoric backgrounds and motivating them to make videos in this space. Participant P11 described her perception of how the country's educational institutions construct a narrative about Bengali people's history:

\begin{quote}\textit{
    From fifth to tenth grade, our textbooks taught us the history of the Indian subcontinent in great detail. By the time we are in tenth grade, we are aware of our history, from ancient times to the history of our liberation war.
}\hfill(P11, female, Bangladesh)\end{quote}

Though our participants' history education has been a strong motivator for their decolonial discourse on YouTube, they complained about cherry-picking in textbooks' historical narratives and thought that their history education should have highlighted a broader set of historical figures. The politics around history education was highlighted strongly by Pakistani YouTubers. For example, P3 explained how textbooks avoid talking about its history with Bangladesh's independence:

\begin{quote}\textit{
    Pakistani textbooks do not have any chapters that mention Bangladesh. The 1947 division of India and Pakistan is mentioned, and all the other things are mentioned as well, but there is nothing mentioned about anything that happened in 1971 [Bangladesh's liberation from Pakistan]. ... The people are unaware of the history.
} (P3, male, Pakistan)\end{quote}

Another Pakistani participant, P13, described how the Pakistani educational system describes the Bengali liberation movement as a result of being ``played by the Indians". This insight aligns with prior works on Pakistani education systems propagating animosity with neighboring countries as a way to strengthen nationalism~\cite{lall2008educate}. Our participants discussed varied selective attention and political efforts to shape accounts of historical events through education systems in different countries, curriculum boards, and partisan administrations. Their perspectives reaffirm Chatterjee's observation of Hindus and Muslims in the subcontinent viewing historical narratives by focusing on different timeframes and historical characters~\cite{chatterjee1993nation}. For example, our Pakistani participants mentioned their education system's exclusive focus on Muslim rule in India while overlooking non-Muslim freedom fighters' contributions. Similarly, our Indian participants expressed concerns about the recent erasures of Muslim historical figures from textbooks under the rise of right-wing politics. Participant P2 expressed disappointment at how historical narratives and communal politics have influenced each other in recent years:

\begin{quote}\textit{
    It is tragic how politics has been shaped. People have been harking back and trying to look for revenge for what happened in the 16th century. ... Maybe they don't blame the British as much as the Mughals, and we know why. The looting of the British was much, much more and much more catastrophic than even the Mughals. But the focus is on the Mughals because that's going to get you ahead in political points.
}\hfill(P2, male, India)\end{quote}

If fragmentary historical narratives institutionalized through education systems continue to condition future generations with exclusionary nationalist views, it would impede the sociopolitical structures of postcolonial nation-states from reforming and reflecting the masses' dreams that served as the basis of anti-colonial movements. Therefore, our participants make videos about regional history to complement and resist institutionally constructed fragmented narratives.

\noindent\paragraph{\textbf{Challenging Transnational Representations in Mainstream Media}}
Our participants from Bangladesh and India, imagining communities based on their shared linguistic identity, as discussed in section~\ref{sec:language}, were motivated to make YouTube content by opportunities to challenge transnational representations of Bengali identity by mainstream media institutions. Indian Bengali YouTubers (e.g., P12, P14, P15) discussed how mainstream media's representation of transnational Bengali culture contributed to fostering and sustaining that sense of affinity in them. They talked about recent collaborations among media personalities, the availability of content from \textit{``Opar Bangla"} on Over-the-top (OTT) platforms, and their contribution to representing cultural practices and underscoring similarities. Participant P12 talked about mainstream media and how it promotes harmony with the neighboring country:

\begin{quote}\textit{
    Indian media talks about Bangladesh. Recently, a singer from Bangladesh came to [a music talent show on Indian Bengali TV] and became the first runner-up. It’s not that we are not in touch with Bangladesh at all.
}\hfill(P12, female, India)\end{quote}

While our Indian participants viewed Bangladeshi culture's representation in their mainstream media as a thread of closeness and imagining a broader Bengali community, our Bangladeshi participants critiqued those as ``caricatures" of and disconnected from contemporary Bangladeshi societies. Thus, our participants were motivated to produce YouTube content that challenged these caricatures and moved towards characterizations that they deemed appropriate and accurate. For example, participants criticized the portrayal of Bangladeshi characters by Indian-Bengali actors in various movies and TV series, as their ways of speaking did not accurately capture any Bangladeshi dialects and accents. Therefore, our participants deem it important to uphold grassroots practices from diverse national contexts instead of curated and stereotyped portrayals.

YouTubers have also discussed how hegemony and prejudices impact transnational media representation. Pakistani YouTubers (e.g., P3, P13) discussed how their mainstream media talks negatively about Indian people, and similarly, Indian YouTubers (e.g., P2, P6) also expressed disappointment about their media vilifying Pakistan. They believe partisan control over mainstream media exacerbates adversarial representations of different countries, religions, and cultural communities, leading to intolerance and communal division within countries and the region. Participant P2 articulated his disappointment about the failure of journalism--gathering, recording, verifying, and reporting information of public importance~\cite{kovach2021elements}, on mainstream media:

\begin{quote}\textit{
    I realize that TV was dying way back in 2010-2011 because there was no growth in television. The respect in TV had gone down. I think Indian TV is talked about enough at this point in time. Even globally, you know how poison-filled and hate-filled it is. ... Pakistan bashing really gets them a lot of content and eyeballs.
}\hfill(P2, male, India)\end{quote}

While our participants mentioned some efforts of institutions like mainstream media to construct a diverse representation of Bengali culture and identity, they critique partisan, communal, and nationalist control by elites like politicians and media personalities, which perpetuates colonial stereotypes and communal division and outweighs its potential to contribute to the decolonial discourse. Despite the inherent politics of online platforms~\cite{gillespie2010politics}, which is out of the scope of this paper, our participants view these platforms as alternative media to build resistance against mainstream media's hegemonic, prejudiced, and adversarial representation. They described using platforms like YouTube as a way to democratize media representation. Instead of seeing people in neighboring countries as adversaries or caricatures through a tainted political and ideological lens of mainstream media, they seek to promote humanizing perspectives to others, for which they strategize video-making on YouTube in ways described in the following section.

\subsubsection{Decolonial Strategies and Reforming ``Outside" Domain of Institutionalization}
Given how institutions are heavily focused on creating and propagating fragmented nationalist narratives as shaped by colonial histories, our informants participated in developing strategies to reform these institutional accounts. In doing so, their decolonial strategies target issues with textbooks and mainstream news production to offer alternative ways for people to discuss complex information and push back against the fragmentary agendas of institutionalized nationalism. These strategies include the creation of (1) political explainers on historical and contemporary geopolitics and (2) journalistic videos about socioeconomic issues at grassroots, national, and subcontinental scales.

\noindent\paragraph{\textbf{Political Explainers to Make Decolonial Discourse Accessible}}
While political discussions on mainstream media are usually not welcoming to general audiences, our participants often structure their videos as political explainers--that explain how something works in a simple and engaging way. As we discussed in the previous subsection, regional geopolitics is one of the central factors of their transnational decolonial discourse. For the decolonial reimagination of sociopolitical structures and institutions, our participants described the accessibility of the general public to political discussions as an imperative factor. To make political discussions accessible, a few of our participants (P2, P4, P7, P8, P11) view these explainers to be an effective means. Participant P2 described his strategies to explain recent political events to his audience:

\begin{quote}\textit{
    We need to make political conversations easier. I think political caricatures really make politics fun, and I think we really need to make politics fun. ... We do political explainers. We do political satire and some humor skits--basically simplifying the news, making it more interesting for the next generation. ... I still remember the massive farm laws protests\footnote{The Indian Farm Bills, aimed to deregulate agriculture and enable direct negotiations between corporations and farmers, sparked widespread protests and was later repealed.~\cite{mishra2021rihanna}.}. ... We used Lego blocks to do to explain the farmer laws. I think that people understood and remembered. So, I think there is a need to ``de-serious-ize" or de-complicate political matters.
}\hfill (P2, male, India)\end{quote}

In their approach to providing more simplified explanations of politics, our participants often focus on controversial topics. They believe that this change in choosing sources of news consumption is driven by the decline in trust in mainstream media like television and the preferences of their primary audience, which primarily consists of young people aged 18 to 35 who seek news and information presented in a conversational manner.

Our participants highlight the historical and contemporary statuses of geopolitical relationships. For example, while Bangladeshi YouTubers speak highly of India's role during the liberation war of 1971, they also highlight current points of contention (e.g., The Teesta and the Ganges water dispute\footnote{Teesta and Ganges are two transboundary rivers with India and Bangladesh in upstream and downstream, respectively~\cite{afroz2013transboundary}.}, killings at border\footnote{The Indian Border Security Force (BSF)'s shoot-to-kill policy at the Bangladesh-India border resulted in the deaths of almost 1,000 people, mostly Bangladeshis, without any prosecutions for these acts of violence~\cite{sarkerindia}.}) in Bangladesh-India diplomatic relationship. In their videos, besides explaining complex diplomatic issues, they imagine better regional geopolitical relationships. Participant P7 underscored the importance of an improved India-Bangladesh relationship, 

\begin{quote}\textit{
    [Attempt to improve Bangladesh's relationship with India] is necessary for our diplomacy, and we cannot avoid it when a large country is on three sides of a smaller country. If you have a neighbor on three sides of your home, will you want to get in conflict with that neighbor? Never. Rather, you would want to maintain a good relationship with them through exchange and equal opportunity. Also, we are India's closest neighbor. So, India would want to be on good terms with us for various reasons.
}\hfill(P7, male, Bangladesh)\end{quote}

Our participants often described political explainers like theirs on YouTube as productive media consumption and better use of this technology. A few of them conveyed this opinion strongly in their videos and motivated their audiences to watch more content of that sort.

\noindent\paragraph{\textbf{YouTube Journalism to Foreground Subaltern Experiences and Perspectives: Challenging Institutionalized Media}}
Some participants described their journalistic endeavors of reporting local incidents and analyzing current affairs as ``YouTube journalism." They particularly focus on localized social and political issues and practices that are often overlooked by mainstream media. They report corruption at local institutions, creating a culture of transparency at the grassroots level. What differentiates our participants' practices from citizen journalism--the collection, dissemination, and analysis of news and information by the general public, especially by means of websites, blogs, and social media~\cite{britannicaCitizenJournalism}, is their exclusive focus on the subaltern (defined in section~\ref{sec:literature_review}), through which they challenge the hegemonic political narratives and help mass people to find their voices in political and economic matters.

A couple of our participants made videos about the rural social stigma and caste-like hierarchy around certain communities and professions that impede the economic progress of those communities. For example, in one video, participant P11 interviewed female garment factory workers. While the issues these workers face are covered by mainstream media periodically after major incidents, she focused on their requirements for safety and rights for fair wages and emphasized how ensuring basic employment benefits for these workers would contribute to the national economy.

Besides identifying challenges in industrial and agricultural sectors, they also identify potentials and propose ways to include underserved communities in the mainstream of economic reformation. For example, Participant P6, an Indian female YouTuber, used the examples of mass mobile app development during the Indian ban on Chinese apps as an example of achieving technological self-sufficiency that leads to unique practices and skills of technicians with no or little educational background in the Global South around repairing devices~\cite{jackson2014learning, irani2019chasing}. Participant P4 calls for attention to these often overlooked local industrial and postcolonial technological potentials:

\begin{quote}\textit{
    If we can utilize those technicians from \textit{Jinjira} (a suburb in Dhaka) in a productive manner, we can design technology on our own terms and according to our own needs. We won't have to depend on others. If those people can develop such complex devices, albeit duplicates, without formal training, why can't they develop similar devices using their own designs? We have to support them financially and legally.
}\hfill(P4, male, Bangladesh)\end{quote}

Here, the participant evaluated the local light engineering talents as indicators of the country's industrial preparedness and possible hubs for sustainable technology and entrepreneurship, which researchers described as crucial for moving toward decolonial economic reformation~\cite{leal2021digital, lu2019feel}.

Toward decolonizing social, political, and economic structures, our participants highlight transferable lessons for Bangladesh, India, and Pakistan due to their similar sociocultural contexts. For example, Pakistani participants emphasized seeking guidance from India and Bangladesh regarding economic development, embracing their approach, and learning from their strategies for economic growth. Similarly, two of our Bangladeshi YouTubers advocated for mass digitization of grassroots economic transactions following India's mass adoption of QR codes for small businesses. In addition to economically underserved communities, Participant P2, an Indian male YouTuber, discussed subaltern communities. like religious minorities, whose protection has been a crucial issue in the region. YouTubers strategically talk about this sensitive issue, where instead of directly pointing out the bad actors (e.g., naming religious fundamentalist entities or individuals), they talk about the systematic problems in similar contexts. Participant P2 explained his approach:

\begin{quote}\textit{
    I made an episode ..., where I talked about the extremism in Bangladesh and how Hindus were being persecuted. The subtle message was that I was basically talking about [persecution of Muslims in] India while I was talking about [violence against Hindus in] Bangladesh. Basically, I was talking about minority persecution.
}\hfill(P2, male, India)\end{quote}

In this case, the participant in his video talked about what led to rising religious extremism in Bangladesh and what steps from the government helped the minorities in crisis. However, since there is limited or politically charged coverage of these kinds of internal and external affairs on mainstream media, our participants use their YouTube channels as platforms for identifying these transferable lessons to dream, commit, and plan actions geared toward better governance.

\section{Discussion}
The double bind of nationalism is evident in its historical contribution to anticolonial movements while simultaneously continuing colonial legacies of division by isolating and fragmenting regional and local identities. This paper focuses on the relationship between nationalism and colonialism and describes YouTubers' motivations and strategies to engage in video-mediated decolonial discourse in transnational contexts. We reported how, through the construction of videos, YouTubers are engaged in decolonial practices that work to revise existing collective understandings and experiences with and of cultural and institutionalized nationalism. In Paul Dourish's seminal work~\cite{dourish2006implications}, he criticized the idea of technological implications as the objective of social computing research. He argued that social computing studies should not approach technologies looking for ``a problem to be eliminated" but how it serves as ``a site for social and cultural production." In that line of social computing research, instead of providing implications for future designs of YouTube or similar platforms, our paper seeks to highlight the participants' creative processes by which they put YouTube into practice and offers an explanation of how this technology becomes a site for social, cultural, and political expression and provides occasions for enacting sociocultural meaning for the Bengali people.

In this discussion, we build on these findings by discussing how technological practices get modularized as driven by sociocultural logic. Building on this, we organize further discussion, first centered around how YouTube's video modality supports decolonial discourse. Then, we reflect on how these decolonial discourses play out beyond the online sphere.

\subsection{Modularization and Polarization in Online Discourse}

Our study leads toward understanding how creative practices around technology embed and perpetuate the understandings and experiences of nationalism of those who use them, which we dub the modularization of identities. In the context of computing, this modularity gives rise to polarization on social media--a phenomenon wherein people are engaged in political discourse only with others who are like them, diverging their political attitudes to ideological extremes~\cite{shi2017cultural, al2021atheists}. In the context of cultural discourse, prior research on representing regional cultural identity using videos described online platforms as the online ``public sphere"~\cite{milliken2008user-canada, milliken2008user-sphere, milliken2008user-youtube}. CSCW and CHI scholarship often build on Habermas' conceptualization of the public sphere to study political deliberation and mitigating polarization~\cite{semaan2014social, nelimarkka2019re}. These works, predominantly conducted in Western contexts, described the population they studied as mostly ``homogeneous"~\cite{semaan2014social, milliken2012older}. While homogeneity aligns with Anderson's ``modular" conceptualization of nations in the Western context, Chatterjee argued that the subcontinental imagination of nations is based on the idea of difference~\cite{chatterjee1993nation}. The stronger the sense of differences within the broader public is, the more likely counterpublics--social groups that develop alternative interpretations of their social identities~\cite{fraser1990rethinking}, are to emerge.

To complement existing work on online cultural and political expression, our paper presents an interesting case from the Global South, where people strongly perceive themselves as members of several intersecting imagined communities, creating a heterogeneous platform identity. We found in our study that YouTubers making videos about Bengali identity and culture subscribe to alternate imagination of their native selves through the lenses of linguistic, religious, and postcolonial states-based nationalism. How these national and broader sociocultural identities are often defined by the exclusion of each other in the local context exacerbates the possibility of counterpublics. How do sociotechnical systems' designs respond to these intersecting imaginations of communities? What does it mean for the decolonial discourse on online platforms?

We know from postcolonial computing scholars that sociotechnical systems are predominately Western-designed products that often strive to support the development of homogeneous communities ~\cite{irani2010postcolonial, dourish2012ubicomp}. These technologies, which modularize people's identities based on culturally sanctioned understandings, encounter complications as they travel and migrate to different cultural contexts. For example, after operating for almost a decade among English-speaking users, the popular Q\&A site Quora expanded to other language-based platforms, with interfaces in local languages but design scaffold not equipped to handle the local cultural nuances. While prior work on Quora exploring identity decolonization found users discussing possibilities of cultural, political, and economic cooperation~\cite{das2022collaborative}, the creation of parallel discursive areas, dubbed ``stages", to address communal and national tension bars participatory and inclusive discourse from developing further on the platform~\cite{das2021jol}. In other words, local communities being balkanized across linguistic, religious, and postcolonial states-based nationalism limits the decolonial discourse at vaguely identifying the objectives and falls short in defining how those objectives could be achieved.

While many social computing studies champion the promise of free participation in online communities, they often overlook or downplay the challenges of intersectional tensions. Our work foregrounds how communications across intersecting national identities or imagined communities (e.g., Pakistani or Indian YouTubers mutually exchanging views, opinions, and ideas with Bangladeshi audiences) within the broader native ethnic identity affect their articulation of social, geopolitical, cultural, and economic objectives for decolonization. 

\subsection{YouTube's Video Modality for Inclusive Decolonial Discourse}
In building on the previous subsection, our work highlighted how YouTubers were engaged in practices that served to perpetuate and support inclusive and participatory discourse. To understand how YouTube, as a discourse platform, supports this diverse exchange of opinions and ideas, we explore the features that support these discourses: (1) length of videos and (2) multimodality.

\subsubsection{Length of Videos for Creating Depth of Perspective} 
Exploring similarities and diversities in Bengali cultural practices across religious and postcolonial boundaries is a major driving force behind our participants' decolonial discourse on YouTube. Though prior works have studied how users represent and archive their cultural practices through short videos on platforms like Douyin and Kuaishou~\cite{lu2019feel, chen2023my}, YouTube's norm centered around making longer videos provides room for the YouTubers to explain their perspectives and unpack nuances and diversities within Bengali culture. For example, our participants make reaction videos where they watch a cultural artifact with their audiences and discuss different aspects of local culture based on what that artifact highlighted. Such informal discussions, in the form of a friendly watch party, resembling \textit{``addas"}~\cite{chakrabarty2008provincializing} function as culturally and politically important platforms for sharing ideas and experiences. Through the discussions on cultural artifacts and documentaries about history, YouTubers emphatically understand and assume others' perspectives--what Habermas dubbed as ``ideal role taking"~\cite{habermas1990moral}. For example, the process of a Pakistani YouTuber making a reaction video after watching a documentary on the Bengali language movement helps that YouTuber to empathize with Bangladeshis, while the Bangladeshi audiences understand an ordinary Pakistani's perspective on their shared history. Compared to short-video sharing platforms, YouTube's norm of longer videos facilitates and encourages such unpacking. Thus, our participants recognize the pluriversality--interconnection among regional experiences and views on historical milestones toward a decolonial imagination.


\subsubsection{Multimodality in Creating Accessibility}
YouTube is primarily a video-based platform. Because videos can combine modes like audio, visual, linguistic, and gestures, they can be categorized as multimodal. Moreover, YouTube supplements videos with textual elements (e.g., titles and descriptions) and spatial-visual elements (e.g., graphical user interface) besides other modes of interaction--making it a multimodal platform. Though digital content of different modalities, such as textual articles and audio podcasts, can be effective in political conversations and media activism online, the multimodality of YouTube allows for more engaging and captivating content. YouTube videos leverage the power of body language, tone of voice, and visuals to enable content creators to communicate with their audiences and effectively convey their message, surpassing the capabilities of other modes of communication. Decolonial discourse through textual communications ~\cite{das2022collaborative, dosono2020decolonizing, das2021jol} can be inaccessible to people who face difficulties with reading and writing in a particular language or, in general. In the postcolonial Indian subcontinent, where literacy rates remain low, YouTube's audio-visual content has the potential to reach broader audiences than textual discussions. Besides, the platform's multimodality can help viewers comprehend YouTubers speaking different languages, particularly when their written text may use different alphabets, but the spoken languages are mutually understandable (e.g., Hindi and Urdu). Thus, YouTube welcomes more diverse people and their perspectives in decolonial discourse.

\subsection{Sociomateriality of Decolonial Discourse Online}
At this stage, we will reflect on the material subjectivities that are shaping these online discourses. How are these decolonial discourses connected to and played out beyond the online sphere? In answering this question, we draw on the concept of sociomateriality, which means that information artifacts (e.g., text, video) are not inherently virtual, but rather, the way the digital world is designed and constructed has material consequences on the physical world, influencing how people act and interact~\cite{bjorn2014sociomaterial, dourish2017stuff}. This dualism of how a community engages with a design is determined by their cultural epistemologies~\cite{irani2010postcolonial}. In turn, that engagement unfolds material consequences into the community~\cite{dourish2017stuff} is explored in the context of identity decolonization by prior work~\cite{das2022collaborative}.

As everything we design, in turn, designs us back~\cite{willis2006ontological}, the creative discursive space that the YouTubers shape shapes their and their audiences' consciousness, which was evident through our participants' encounters in a foreign hotel or them receiving invitations through newly formed social relationships on YouTube based on linguistic similarities. Moreover, in the contemporary landscape of decolonization, scholars find technology to be more effective in executing decolonial actions and generative of real-world change~\cite{laenui2000processes}. Our participants' journalistic efforts to identify local economic and industrial potential and report corruption resemble recent work on online activism leading toward promoting transparency and causing policy changes within governments and organizations~\cite{hansson2021organizing, livingstone2021make}.

\section{Conclusion: Invitation to Critically Study Nationalism in CSCW}
Our study presents a nuanced account of how coloniality has shaped marginalized communities' ideas and experiences of nationalism. We describe the ways in which individuals' conceptualization of their collective identities across various dimensions, along with their discernment of the nationalism perpetuated by different institutions, motivate their active participation in video-mediated decolonial discourse. We also highlight their decolonial strategies realized through social interviews, decolonial storytelling, political explainers, and YouTube journalism to foster interconnected cultural understanding, construct comprehensive historical narratives, and foreground local and grassroots perspectives, which aim to strengthen and reaffirm native identities on social-psychological levels, as well as facilitate reformations in sociopolitical and economic spheres. While this paper focuses on YouTubers' motivations and strategies for video-mediated discourse, in our future work, we will look into the challenges these YouTubers face as content creators from the Global South and how the platform both supports and impedes their expression of cultural identity and participation in sociopolitical discussions.

Finally, we develop a call to action for the critical study of nationalism in social computing research. Although nationalism plays a significant role in shaping people's collective identity, there has been a lack of critical examination of this topic within the field of social computing and HCI by identity scholars. Among social computing scholarship that has foregrounded the idea of nationalism, most have used nationalism as a typology in computational social science research to study topical focuses of political campaigns~\cite{panda2020topical, schafer2017japan} and hate speech (e.g., xenophobia, anti-Semitism, racism)~\cite{olteanu2018effect, zannettou2020quantitative, nagar2021empirical} in social media. However, such use of nationalism as a classification in datasets does not speak to its complex and deep entanglements with social relationships and how those translate through different institutions and technologies. While some scholars have recently looked at how sociotechnical systems' designs reinforce platformed racism and how it relates to white nationalism and supremacy~\cite{wu2022conversations, lalone2014values, hagen2019emoji}, to the best of our knowledge, our paper delves into the complexity of nationalism to a deeper extent previously unexplored in CSCW and social computing scholarship. We argue that there is a pressing need for further research on nationalism to deepen our understanding of its evolving manifestations, impacts, and interconnections with various socio-political dynamics. By exploring its complexities across diverse geopolitical and cultural contexts and time periods, we can examine the multifaceted nature of nationalism and its role in shaping identities and interaction in online communities, thus offering valuable insights for informed design, content moderation, and platform governance.

\section{Acknowledgements}
The research was partially funded by NSF grant no. CRII-1657429. The authors would also like to thank the members of the ADA lab at the University of Colorado Boulder for their support and feedback. We also thank Lilly Irani for her insights during the proposal phase of this paper.
\bibliographystyle{ACM-Reference-Format}
\bibliography{sample-base}

\appendix
\section*{Appendix}
\section{Geographic Composition of the Indian Subcontinent}\label{app:indian_subcontinent}
Scholars have different opinions about the geographic demarcation of the Indian subcontinent~\cite{mcleod2015history, bhopal2007ethnicity}. Scholars agree that the region consists ``at least of India, Pakistan, and Bangladesh"~\cite{zeidan2022indian}. However, in that case, the term becomes subject to the politics of nomenclature. While some sources describe the region using the phrase ``Greater India"~\cite{vivekanandan2018indianisation}, Pakistani national historiography often champions the phrase ``Indo-Pakistan subcontinent"~\cite{mann2014south}. Again, scholars who use the term ``South Asia" interchangeably with the term ``Indian subcontinent" view Afghanistan, Bhutan, Nepal, Sri Lanka, and the Maldives as parts of this region~\cite{norwine1988third}. However, some scholars have argued against such views. According to them, Afghanistan is a central Asian country~\cite{masson1992history, mcleod2015history}, and while Sri Lanka and the Maldives have cultural similarities with the countries in the Indian subcontinent, these island countries lack geographic contiguity with the region~\cite{mcleod2015history}. In the case of Bhutan and Nepal, these countries used to be British protectorates and not British colonies~\cite{onley2009raj}. Therefore, in our study, we focused on Bangladesh, India, and Pakistan--three countries that, scholars unanimously agree, are parts of the Indian subcontinent and were subjugated together under British colonial rule till 1947.

\section{Local and Indigenous Identities in the Indian Subcontinent}\label{app:local_and_indigenous}
Ways to conceptualize indigeneity and identify indigenous people vary across different regions and are subject to emancipatory politics~\cite{baviskar2006politics, eubanks2018we}. For example, in the Indian subcontinent, a region with long history of human migration, though ``indigenous" is usually used to identify the tribal \textit{adivasi} groups (e.g., Bhil, Munda, Santhal, etc.)~\cite{britannicaAdivasiIndigenous}, non-tribal groups like the Bengalis, Gujratis, and Oriyas, also have a long history of settlement~\cite{xaxa1999tribes}. To inclusively recognize these communities who lived in the region since long before the British colonization, following prior work~\cite{das2022collaborative}, we are using the phrase ``local and indigenous" communities.

\end{document}